\documentclass[11pt,a4paper]{article}
\usepackage{jheppub}
\pdfoutput=1

\usepackage{wasysym}

\usepackage{amsmath}
\usepackage{amssymb}  % checkmark
\usepackage{mathrsfs} % mathscr
\usepackage{graphicx}
\usepackage{dcolumn}
\usepackage{xcolor}
\usepackage{url}
\usepackage{rotating}
\usepackage{multirow}
\usepackage{ulem}

\usepackage{bbold}  % \mathbb{1} for the identity matrix

\newcommand\identity{1\kern-0.25em\text{l}}
\def\gsim{\raise0.3ex\hbox{$\;>$\kern-0.75em\raise-1.1ex\hbox{$\sim\;$}}}
\def\lsim{\raise0.3ex\hbox{$\;<$\kern-0.75em\raise-1.1ex\hbox{$\sim\;$}}}

\usepackage{comment}
\usepackage{subcaption}

\title{
\vspace*{2cm}
Long-Lived HNLs via ALP Portal at the LHC
}

\author[a,b]{Rebeca Beltr\'an,}
\author[a]{Chandan Hati,}
\author[a]{Martin Hirsch,}
\author[a]{Ana Martín-Galán}

\emailAdd{beltran-lloria.rebeca@ucy.ac.cy}
\emailAdd{chandan@ific.uv.es}
\emailAdd{mahirsch@ific.uv.es}
\emailAdd{Ana.Martin@ific.uv.es}

\affiliation[a]{Instituto de F\'{\i}sica Corpuscular  (CSIC-Universitat de València),  46980 Paterna, Spain}
\affiliation[b]{Department of Physics, University of Cyprus, 1678 Nicosia, Cyprus}

\abstract{

  Heavy neutral leptons (HNLs) and axion-like particles (ALPs) are
  both considered well-motivated candidates for beyond the standard
  model (BSM) physics. If ALPs with sizable couplings to gluons
    exist, they will be abundantly produced at the LHC. Therefore,
  HNLs produced via the ALP portal may provide unprecedented
  sensitivities to HNL parameters. Here, we study the prospects for
  the high-luminosity LHC to search for long-lived HNLs. We consider
  future far detectors as well as ATLAS in our simulations. In the
  limit where the ALP mass is above the TeV scale, HNLs are
  effectively produced by a dimension-8 operator connecting HNL pairs
  to gluons. For completeness, we therefore also calculate future LHC
  sensitivities for HNLs produced via $N_R$SMEFT operators with
  gluons.
 
}

\makeatletter
\gdef\@fpheader{\phantom{a}}
\makeatother
           
\begin{document}
\maketitle

% !TEX root = ../NRALP.tex
\section{Introduction\label{sec:intro}}

The advent of the long-lived particle programme at the CERN hadron
collider~\cite{Alimena:2019zri, Lee:2018pag} has rekindled interest in heavy
neutral leptons (HNLs) also from the theoretical side. Minimal HNLs
are heavy neutral leptons that interact only via charged and neutral
current interactions, suppressed by some small mixing angle, with
standard model (SM) particles.  Sensitivity estimates for minimal
HNLs for the high-luminosity phase of the LHC have been published
for ATLAS and CMS~\cite{Cottin:2018kmq, Cottin:2018nms,
  Drewes:2019fou, Bondarenko:2019tss,Liu:2019ayx} and for proposed
``far detectors'' \cite{Helo:2018qej,Hirsch:2020klk,Curtin:2018mvb},
such as MATHUSLA~\cite{Chou:2016lxi}, CODEX-b~\cite{Gligorov:2017nwh},
FASER~\cite{Feng:2017uoz}, ANUBIS \cite{Bauer:2019vqk} or MAPP and MAPP2 \cite{Pinfold:2019nqj,Pinfold:2019zwp}.  A summary
of the current bounds for minimal HNL models can be found, for
instance, in~\cite{Bolton:2019pcu,Bolton:2022pyf}, and for the most recent bounds from the LHC experiments, see \cite{ATLAS:2025qbs}. Also, sensitivity projections for probing minimal HNLs at future experiments can be found in \cite{deBlas:2025gyz} and for a comprehensive review discussing various aspects of HNLs, see e.g. Ref.~\cite{Abdullahi:2022jlv}.

However, in many UV complete models, HNLs are not the only BSM
ingredients. Such models can include additional new particles, such as
$Z'$ bosons~\cite{Deppisch:2019kvs,Chiang:2019ajm} or
leptoquarks~\cite{Dorsner:2016wpm,Cottin:2021tfo}, for instance.  In
all these examples, the additional new particles must be relatively
heavy, roughly of the order TeV or above. Since none of these exotics
have so far been seen in LHC direct resonance searches, the
Effective Field Theory (EFT) approach provides a robust
model-independent framework to explore such scenarios systematically.
Standard model effective field theory with right-handed neutrinos,
$N_R$SMEFT, has been studied in a number of works, both from more
theoretical as well as phenomenological
sides~\cite{Bell:2005kz,Graesser:2007yj,
  Graesser:2007pc,delAguila:2008ir, Aparici:2009fh,Liao:2016qyd,
  Beltran:2023ymm,Beltran:2024twr}.  Predictions for dimension-6
$N_R$SMEFT operators at the high-luminosity LHC have been explored, for
example, in~\cite{Cottin:2021lzz,Beltran:2021hpq,Beltran:2025ilg}. A
compilation of various constraints on these operators from a wide
range of HNL searches has been found, for instance,
in~\cite{Fernandez-Martinez:2023phj}.

HNLs, on the other hand, are not the only BSM particles that could be
light enough to be produced on-shell at the LHC.  Axion-like particles
(ALPs) are pseudoscalars that can arise as pseudo-Goldstone bosons of
a spontaneously broken global symmetry~\cite{Jaeckel:2010ni}. The
couplings of the ALP with the SM fields are protected by an
approximate classical shift symmetry. However, different from the
classical axion~\cite{Peccei:1977hh,Peccei:1977ur, Weinberg:1977ma,
  Wilczek:1977pj, Dine:1981rt,Kim:1979if}, the mass of the ALP is a
free parameter, $m_a$, and can take much higher values.  ALPs have
been investigated across a variety of contexts, including flavor
physics~\cite{Bauer:2021mvw}, direct collider
searches~\cite{Bauer:2017ris,Brivio:2017ije,Biekotter:2025fll},
long-lived signatures, and/or in connection with dark
matter~\cite{Alimena:2019zri,Jaeckel:2010ni,Alekhin:2015byh}.

ALPs do not have renormalizable couplings to SM fields. However, at
$d=5$ ALPs can couple to fermions as well as to all SM gauge
fields. At LHC energies, the gluon parton luminosities are large and
thus, ALPs coupled to gluons could be copiously produced. ALPs can
decay to HNLs as well, and it has been
shown~\cite{deGiorgi:2022oks,Marcos:2024yfm} that in such a BSM
extension ALPs can act as a very efficient portal for producing
HNLs. The production cross sections for pair production of HNLs $pp
\to a^{*} \to N N$ have been calculated
in~\cite{deGiorgi:2022oks,Marcos:2024yfm} to estimate sensitivities
for the new physics scale at the high-luminosity LHC.
However, these studies did not address the subject exhaustively. 
In particular, it is assumed in \cite{deGiorgi:2022oks,Marcos:2024yfm}
that the HNLs decay promptly into $N \to W^{\pm} l^{\mp} \to jj l$
and that $m_a \simeq 2$ GeV $\ll m_N$.  Our current paper is meant to
fill several gaps left by these earlier works~\cite{deGiorgi:2022oks,
  Marcos:2024yfm}: (i) we explore heavier ALP masses $m_a > 10$
GeV; (ii) we simulate the long-lived HNL decays for both ATLAS and the
far detectors; (iii) we perform an analysis for the case where HNLs
couple directly to gluons via effective $d=7$ and $d=8$ operators. We
note that two recent studies~\cite{Wang:2024mrc, Wang:2024prt} explored a similar idea as this work. However, they focused on ALPs produced in
meson decays (therefore limited to ALP masses $m_a \lsim 5$ GeV) and derived sensitivity prospects for SHiP and Belle, providing a nice complement to our work.

The rest of this paper is organized as follows. In section
\ref{sec:setup}, we discuss the theoretical basis. The ALP and HNL
Lagrangians are given, and effective operators for $N_R$SMEFT that
generate couplings of the HNL to gluons are defined. We also discuss
potential UV complete model examples where these effective
interactions with enhanced couplings of ALPs to gluons and HNLs can be
realized.  Section \ref{sec:pheno} presents numerical results. In
section \ref{subsec:sim}, we briefly discuss our simulation setup. In
subsection \ref{subsec:prod}, we discuss production cross sections and
ALP decays.  Numerical results for future sensitivities are then given
in subsection \ref{subsec:res}.  We then close with a short discussion
and summary. In the appendices, we discuss current limits on the ALP
coupling to gluons from LHC dijet searches (appendix~\ref{sec:app}) and compare sensitivities of the most recent to older configurations of the MATHUSLA and ANUBIS 
detectors (appendix~\ref{sec:appB}).

% !TEX root = ../NRALP.tex
\section{Theoretical setup: HNLs and ALPs\label{sec:setup}}

\subsection{Minimal HNLs\label{subsec:HNL}}

A heavy neutral lepton (HNL) is defined by its charged and neutral
current interactions with the SM leptons
\begin{eqnarray}\label{eq:CCNC}
{\cal L} &=& \frac{g}{\sqrt{2}}\,  \sum_{\alpha,j}
 V_{\alpha N_j} \bar l_{\alpha,L} \gamma^{\mu} N_{j} W^-_{L \mu} 
 +\frac{g}{2 \cos\theta_W}\ \sum_{\alpha, i, j}V^{L}_{\alpha i}
 V_{\alpha N_j}^*  \overline{N_{j}} \gamma^{\mu} \nu_{i,L} Z_{\mu}
     \hskip2mm + {\rm h.c.}
\end{eqnarray}
where $\alpha=e,\mu,\tau$, $i=1,2,3$ for the SM generations and mass
eigenstates and $j$ runs over the number of HNLs. For consistency, one
also has to add an interaction term among the SM Higgs, the SM
neutrinos, and the HNL to eq.~\eqref{eq:CCNC}.  This Lagrangian does
not specify the nature of the HNL, which could be either Majorana or
Dirac.  Experimental searches will then provide constraints on mixing
parameters, $V_{\alpha N_j}$, versus HNL mass. We refer to HNLs with
only the interactions in eq.~\eqref{eq:CCNC} as minimal HNLs.

The Lagrangian in eq.~\eqref{eq:CCNC} does not relate the mixing
parameters $V_{\alpha N_j}$ to active neutrino masses (and mixings) as
measured in oscillation experiments. In order to do so, one needs
to specify the underlying neutrino mass model, which could be any
variant of the seesaw: (i) a classical type-I seesaw
\cite{Minkowski:1977sc,Yanagida:1979as,Mohapatra:1979ia,GellMann:1980vs};
(ii) the inverse seesaw \cite{Mohapatra:1986bd}; (iii) a linear
seesaw \cite{Akhmedov:1995ip,Akhmedov:1995vm} or any other valid model. 
The simplest variant is the type-I seesaw, which adds right-handed 
neutrinos to the SM with Yukawa couplings to SM the leptons and
Majorana mass terms
\begin{equation}\label{eq:Yuk}
  {\cal L}^{Y} = Y^{\nu,*}_{ij}
  \left(\overline{N_{R_j}}L_i^{\alpha}\epsilon_{\alpha\beta} \right)H^{\beta}
  + \frac{1}{2}M_{M,jj}\overline{N_{R_j}^c}N_{R_j}
  +\hskip2mm {\rm h.c.}
\end{equation}
where $\epsilon_{\alpha\beta}$ denotes the $SU(2)$ contraction
  and $M_M$ can be chosen to be diagonal without loss of
generality. We have added a subscript $M$ to indicate the Majorana
nature. The HNL mass will be denoted by $m_N$.  After electroweak
symmetry breaking, this Lagrangian will generate to leading order
the neutral and charged current interactions of
eq.~\eqref{eq:CCNC}. At least two copies of $N_R$ are needed in this
model, in order to be able to fit experimental neutrino oscillation
data and, in this simple model, $V_{\alpha N_j}$ can be related to
light neutrino masses and mixings via the Casas-Ibarra parametrization
\cite{Casas:2001sr}.  In this work, we will not be interested in
performing a detailed fit of active neutrino data. Since the relation
between neutrino masses and the mixing angles $V_{\alpha N_j}$
strongly depends on the choice of neutrino mass model, we will simply
treat the entries in the mixing matrix, $V_{\alpha N_j}$, as free
parameters in our numerical scans.  We will, however, assume that HNLs
are Majorana particles. Results for Dirac HNLs will be similar.

\subsection{The ALP Lagrangian at $d=5$\label{subsec:ALP}}

ALPs are pseudoscalars with couplings to SM fields that are protected
by an approximate shift symmetry.  The shift symmetry is broken only
by the presence of a mass term, $m_a$. The ALP lagrangian up to $d=5$
can be written as \cite{Georgi:1986df}
\begin{eqnarray}\label{eq:lag}
  {\cal L}_a &=& \frac{1}{2}\partial_{\mu}a\partial^{\mu}a 
             - \frac{1}{2} m_a^2 a^2 \\ \nonumber
          &-&\sum_{X} \frac{c_{X{\tilde X}a}}{\Lambda} a X^{\mu\nu}{\tilde X}_{\mu\nu}
             -\sum_{\psi} \frac{c_{\psi \psi a}}{\Lambda}\partial_{\mu}a
             ({\overline\psi}\gamma^{\mu}\psi)
             - \frac{c_H}{\Lambda} \partial_{\mu}a  
    \big( H^\dagger i \overleftrightarrow{D}^\mu\,H \big),
\end{eqnarray}
where $X^{\mu\nu}$ represents a SM field strength tensor, with
$X=B,W,G$, and ${\tilde X}_{\mu\nu}$ is its dual. $\psi$ denotes a
generic SM fermion, and in our case also includes the right-handed
neutrino, $N_R$.  We have written down also a possible coupling of the
ALP to the SM Higgs for completeness, although we will not study this
term. For the purposes of this work, the most relevant ALP couplings
are $c_{G{\tilde G}a}$ and $c_{NNa}$.

\subsection{Effective operators involving HNLs and gluons\label{subsec:EFT}}

At the LHC, ALPs can be produced on-shell from their coupling to
gluons up to masses roughly between $m_a =(1-2)$ TeV. For larger
masses, the ALP can be integrated out from the Lagrangian
eq.~\eqref{eq:lag} and higher-dimensional SMEFT (and, in our case,
$N_R$SMEFT) operators appear at $d=8$, of the form
\begin{equation}\label{eq:d8}
  {\cal L}_{\rm eff} = \frac{c_{X\psi}}{\Lambda^4} \sum_{X,\psi}
     X^{\mu\nu}{\tilde X}_{\mu\nu}({\overline\psi}\partial_{\rho}\gamma^{\rho}\psi).
\end{equation}
Here, in the spirit of effective field theory, we have set
$m_a=\Lambda$, which of course will not be true in an explicit ALP
model. Note, in particular, there is an effective coupling of two
gluons to a pair of $N_R$ in eq.~\eqref{eq:d8}.

There are, however, effective operators that couple right-handed
neutrinos to gluons at a lower dimension. These appear already at
  $d=7$ and we therefore include them for consistency. They can be
written as \cite{Liao:2016qyd}
\begin{equation}\label{eq:d7}
  {\cal O}^{d=7}_{GN} = \frac{c_{GN}}{\Lambda^3} 
          G^{\mu\nu}G_{\mu\nu}\overline{N_R^c}N_R +
                 \frac{c_{{\tilde G}N}}{\Lambda^3} 
  G^{\mu\nu}{\tilde G}_{\mu\nu}\overline{N_R^c}N_R\,.
\end{equation}
Note that these operators violate lepton number by two units.

\subsection{Comments on UV Complete Model Realizations}{\label{sec:models}}
In this work, we are interested in cases where the ALPs can present enhanced couplings to both gluons and HNLs. While the coupling of neutrinos to an ALP will induce a coupling of the ALP with gluons at the two-loop level (via mixing of the ALP with the
$Z$ boson), such couplings would usually be very suppressed. In what follows, we comment on some minimal UV completions that can realize sizable ALP couplings to both gluons and HNLs.

If the lepton number is associated with a global $U(1)_L$ symmetry, the spontaneous breaking of such a symmetry by a scalar $S=\frac{f_N +\hat{s}}{\sqrt{2}} e^{i J/f_N}$ can be related to a pseudo-Nambu-Goldstone boson (PNGB) called Majoron~\cite{Chikashige:1980ui,Schechter:1981cv,Garcia-Cely:2017oco}. The Majoron $J$ can become a massive ALP through connection with
gravity~\cite{Akhmedov:1992hi,Rothstein:1992rh,Alonso:2017avz}, or
simply by the presence of a soft global symmetry-breaking term in the
Lagrangian~\cite{Gu:2010ys,Frigerio:2011in}, making its mass
potentially a free parameter. This setup presents a tree-level coupling of the ALP with lepton-number-carrying HNLs, which is inversely proportional to $f_N$~\cite{Garcia-Cely:2017oco}. To realize an enhanced coupling to gluons simultaneously, additional ingredients are
necessary. A minimal addition to the above construction can be to add
a color multiplet charged under the same $U(1)_L$, e.g., a fermion
$\Psi(8,1,0,1)$, where the quantum numbers in the brackets denote
the representations under ($SU(3)_C$, $SU(2)_L$, $U(1)_Y$,
$U(1)_L$). This will lead to a coupling of the ALP to gluons
via the triangle anomaly of the form
\begin{equation}{\label{Lag:gluon-maj}}
  \mathcal{L}\supset -3 \frac{n_\Psi \alpha_s}{8\pi}
   \frac{J}{f_N} \tilde{G}^{A}_{\mu\nu}{G}^{A\mu\nu}\, ,
\end{equation}
where $n_\Psi$ denotes the number of generations of $\Psi$. To ensure that the ALP couplings to neutrinos and gluons are sizable, $f_N$ must be small, which can be realized, for instance, in an inverse seesaw setup~\cite{Bansal:2022zpi}.

Another straightforward possible setup to have enhanced couplings to both
HNLs and gluons would be to add vector-like quarks ($\psi_{q_{L,R}}$)
and vector-like SM singlet HNLs ($N_{L,R}$) to the SM. One can then charge them under
a $U(1)$ global symmetry, such that the terms
\begin{equation}{\label{Lag:gluon-maj2}}
  \mathcal{L}\supset Y_{N} \left( \overline{N_L} N_R \right) S + Y_{\psi_q} \left(\overline{\psi_{q_L}}\psi_{q_R} \right) S\, 
\end{equation}
are allowed (instead of bare vector-like mass terms), and as in the
previous model $S=\frac{f_a +\hat{s}}{\sqrt{2}} e^{ia/f_a}$. Such a
construction will lead to one-loop couplings with gluons via the
triangle anomaly of heavy vector-like quarks, similar to the previous
construction. It will allow for a tree-level coupling of the ALP $a$
to the HNLs. Dirac masses for light neutrinos can be easily
generated in such a case via a radiative mechanism by adding
extra leptoquark states, see e.g. Ref.~\cite{Batra:2025gzy} or
via a tree-level Dirac seesaw by introducing additional Higgses and
light right-handed partners for the SM left-handed neutrinos.

Finally, an ALP with a significantly enhanced coupling to gluons can also be naturally realized in several possible scenarios proposed in the literature to solve the axion quality problem~\cite{Kamionkowski:1992mf,Barr:1992qq,Ghigna:1992iv,Holman:1992us}. One such realization proposes introducing a mirror copy of the SM, in
which both sectors, including the $\bar{\theta}$ parameter, are
assumed to be symmetric under a $\mathbb{Z}_2$
symmetry~\cite{Hook:2019qoh}. In such a construction, the
$\mathbb{Z}_2$ symmetry is softly broken by the vacuum expectation
value of the mirror Higgs, which is assumed to be much larger than the
SM Higgs, leading to a heavier mass spectrum for the mirror
quarks. This leads to a larger confinement scale in the mirror QCD
sector, potentially resulting in a significant enhancement of the
axion mass, making TeV-scale axions with a large coupling to gluons
viable. Some alternative constructions embed the SM QCD gauge group
into larger semi-simple groups~\cite{Agrawal:2017ksf,Gaillard:2018xgk,Csaki:2019vte} or modify the UV running of the QCD structure constant using extra-dimensional
frameworks~\cite{Flynn:1987rs,Holdom:1982ex,Holdom:1985vx} to realize
heavy axions with enhanced coupling to gluons.

% !TEX root = ../NRALP.tex
\section{Phenomenology\label{sec:pheno}}

\subsection{Simulation setup \label{subsec:sim}}

We have generated \texttt{UFO} models \cite{Degrande:2011ua} for the
Lagrangians discussed in section \ref{sec:setup} above using
\texttt{FeynRules} \cite{Alloul:2013bka}. Production cross sections
and decay widths for the ALP have then been calculated using
\texttt{MadGraph5} \cite{Alwall:2011uj,Alwall:2014hca}.
(We use  \texttt{MG5$\_$aMC$\_$v3.5.9} in the numerical simulations.)
The right-handed neutrino will decay via mixing to SM particles and,
  for simplicity, in our simulations we assume it mixes exclusively
  with electron neutrinos, i.e. $V_{\alpha N} = V_{eN}$.\footnote{Based on a previous study \cite{Beltran:2021hpq}, we
  expect that results for muons will be very similar, while taus will
  offer a poorer sensitivity owing to their reduced reconstruction
  efficiencies at the detectors.} The corresponding decay widths
have been calculated using the analytical formulas given in the
literature \cite{Bondarenko:2018ptm}. In this scenario,
  electroweak scale HNLs with suppressed mixings are expected to be
  long-lived enough to leave displaced signatures in a main LHC
  detector or in a planned far detector experiment. For instance, $m_N
  = 50$~GeV and $|V_{eN}|^2 = 10^{-12}$ lead to $(c\tau)_N \approx 1$~m, while for $m_N
  = 500$~GeV and $|V_{eN}|^2 = 10^{-15}$, the proper decay length is $(c\tau)_N \approx 1$~mm.\footnote{We recall that the decay width (the inverse of the proper decay length) scales with $m_N^5$ for $m_N< m_W$, and for $m_N > m_W$, it transitions to being proportional to $m_N$ as 2-body decays dominate the width. In addition, the decay width scales with $|V_{\ell N}|^2$ for all HNL masses.} Below, we describe the simulation procedure for both the far detectors and
  ATLAS as one of the main LHC detectors. We do not
  simulate explicitly the CMS detector, for which we expect
  similar results.

For the simulation of the far detectors we use the Displaced Decay
Counter (DDC) \cite{Domingo:2023dew}, with some modifications relative
to the detectors described in the original publication
\cite{Wang:2025esc}. These updates concern essentially two
experiments. First, the MATHUSLA detector, originally proposed in
\cite{Chou:2016lxi}, has gone through various design iterations. In
the most recent design \cite{MATHUSLA:2025zyt}, the detector is
proposed to have dimensions of $40\times 40\times 16$ m$^3$. This
latest version has been recently included into the DDC simulation
software \cite{Domingo:2023dew}. We will refer to this setup as
MATHUSLA-40 in the following.  Second, ANUBIS \cite{Bauer:2019vqk} was
originally proposed to be installed in a service shaft above the ATLAS
detector. However, recently the collaboration has discussed an updated
design. The updated configuration \cite{Brandt:2025fdj} plans to
install the detector components directly onto the ceiling of the ATLAS
cavern. This geometry is referred to as ANUBIS-C throughout this
paper. Also this change has been now included into DDC
\cite{Wang:2025esc}. In appendix \ref{sec:appB}, we compare how these
changes in design affect the expected sensitivities of the two
detectors.

For the far detectors, the DDC just calculates the total number of
events decaying in each of the implemented detector volumes, without
including efficiencies for event detection. This is equivalent to
assuming efficiencies of 100\%, which is certainly too optimistic.
However, more important for the sensitivity estimate is the assumption
that the far detectors will be background-free experiments. This will,
most likely, not be the case for any of the proposed detectors.
However, so far, only ANUBIS has published an estimate for the
expected number of background events \cite{Brandt:2025fdj}. Very
  recently, ANUBIS \cite{ANUBIS:2025sgg} published an update of their
  background estimate.\footnote{The latest ANUBIS paper
    \cite{ANUBIS:2025sgg} appeared on the \texttt{arXiv} the same day
    as v1 of our current work.}  According to \cite{ANUBIS:2025sgg},
  ANUBIS-C is expected to have up to $182.4 \pm 12.2$ background
  events in ${\cal L}=3$ ab$^{-1}$. We therefore decided to show
  estimated sensitivity lines for ANUBIS-C for 28 events (roughly a
  2$\sigma$ upper bound for 195 background events) and for 195 events
  (as used by the collaboration in \cite{ANUBIS:2025sgg}). For all
other experiments we will show the 3 (and 30) event contours.

For the simulation of ATLAS we use custom made software. This part of the calculation follows the description given in previous publications
on HNL searches \cite{Cottin:2018kmq, Cottin:2021lzz,Cheung:2024qve,
  Beltran:2025ilg}. In particular, we follow the strategy used in
\cite{Cottin:2021lzz} for the displaced vertex (DV) reconstruction, which targets long-lived HNLs decaying as $N \to e jj$. Events are required to contain an electron with $p_T^e > 120~\mathrm{GeV}$ and $|\eta_e| < 2.47$. The displaced decay is required to occur within the ATLAS inner tracker volume, imposing $4~\mathrm{mm} < r_{\mathrm{DV}} < 300~\mathrm{mm}$ and $|z_{\mathrm{DV}}| < 300~\mathrm{mm}$. A DV is reconstructed from at least four displaced charged particle tracks with transverse impact parameter $|d_0| > 2~\mathrm{mm}$, one of which must be associated with the trigger electron. For further details of the reconstruction and event selection, see \cite{Cottin:2021lzz}, where also plots of the detection efficiencies as a function of the mass $m_N$ can be found.
Since our right-handed neutrinos are pair produced, the final efficiency for detection contains an additional factor of 2 per event.

\subsection{Production cross sections and ALP decays\label{subsec:prod}}

\begin{figure}[t]
    \centering
    \hspace{5mm}
    \includegraphics[scale=0.9]{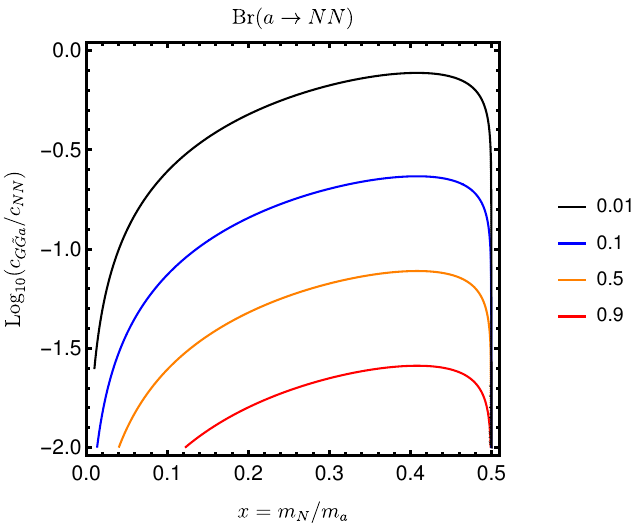}
    \caption{Contours of branching ratios $\text{Br}(a\to NN)$ as a function of Log$_{10}(c_{G{\tilde G}a}/c_{NNa})$ and
      $x=m_N/m_a$. All other Wilson coefficients are assumed to be
      zero in this plot.}
    \label{fig:BrNN}
\end{figure}

Due to their derivative couplings, ALP decays to fermions are
proportional to $\Gamma(a\to \overline{\psi}\psi) \propto m_a
m_{\psi}^2$, whereas ALP decays to massless gluons (and photons) are
proportional to $\Gamma(a\to gg/\gamma\gamma) \propto m_a^3$. If all
Wilson coefficients, $c_i$, are of the same order, a heavy ALP ($m_a \gg m_\psi$) therefore preferentially decays to gluons (jets). The second most
important decay mode, however, will be to pairs of $N_R$, if the $N_R$
mass is larger than the mass of any SM fermion with a non-zero Wilson
coefficient. In our numerical simulations we will therefore put only
two Wilson coefficients non-zero, $c_{G{\tilde G}a}$ and
$c_{NNa}$. Unless there is a strong hierarchy in the Wilson
coefficients and $c_{G{\tilde G}a}$ and $c_{NNa}$ are much smaller
than all others this should be a good approximation.

Fig.~\ref{fig:BrNN} shows an example for the branching ratio Br$(a
\to NN$) in a contour plot as a function of $x=m_N/m_a$ and
$c_{G{\tilde G}a}$. The maximum branching ratio is reached at $x
\simeq 0.4$. For $c_{G{\tilde G}a}=c_{NNa}$ only about $\sim 0.5$\%
of the ALPs will decay to $NN$. However, $c_{G{\tilde G}a}$ could be
naturally much smaller than $c_{NNa}$. For example, one expects that
in UV models for the ALP $c_{G{\tilde G}a}$ will be generated
effectively at one-loop order, whereas $c_{NNa}$ could be tree-level
generated.  It it thus not unnatural to expect a certain hierarchy
between $c_{G{\tilde G}a}$ and $c_{NNa}$. Fig.~\ref{fig:BrNN}
therefore shows branching ratio contours down to $c_{G{\tilde G}a}/
c_{NNa} = 10^{-2}$.  Below $c_{G{\tilde G}a}/c_{NNa}=0.05$, and
depending on the $N$ mass, ALPs can decay dominantly to two $N$.
The ratio $c_{G{\tilde G}a}/c_{NNa}$ strongly influences the $N$
production cross section, as we will discuss next.

\begin{figure}[t]
    \centering
    \hspace{2cm}
    \includegraphics[scale=0.9]{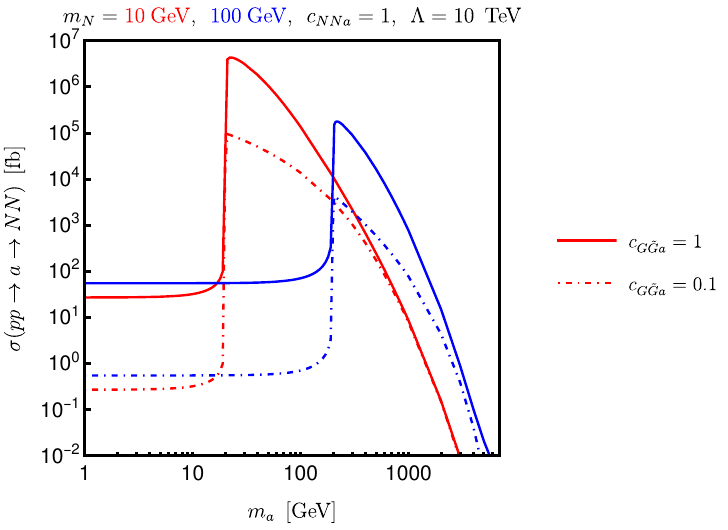}
    \caption{Cross section $\sigma(pp \to a^{(*)} \to NN)$ in fb for
      $\sqrt{s}=14$ TeV as a function of the ALP mass, $m_a$, for two
      different choices of $m_N$ and $c_{G{\tilde G}a}$. There are
      three different kinematic regimes: (i) $m_a \le 2 m_N$; (ii) $2
      m_N \le m_a \lsim 1$ TeV and (iii) $m_a \gsim $ TeV. In regime
      (ii) the ALP can be on-shell and the cross section is enhanced
      due to the small width of the ALP. In regime (iii) the ALP is
      too heavy to be produced on-shell and the cross section reduces
      to the contribution of an effective operator coupling HNLs to
      gluons, see text. }
    \label{fig:xsec}
\end{figure}

Fig.~\ref{fig:xsec} shows some example cross sections for $\sigma(pp
\to a^{(*)} \to NN)$ at the LHC as function of the ALP mass, $m_a$,
for two choices of $m_N$ and $c_{G{\tilde G}a}$.  For this plot, we
have fixed $c_{NNa}=1$ and $\Lambda=10$ TeV.  The latter is motivated
by the current limit on $c_{G{\tilde G}a}/\Lambda$ from dijet searches
at the LHC \cite{CMS:2019gwf,ATLAS:2018qto}, which we discuss in
appendix \ref{sec:app}. We stress, however, that there are currently
no limits on $c_{G{\tilde G}a}/\Lambda$ from dijet searches at the LHC
for ALP masses below $m_a=450$ GeV. The maximum of the cross
  section is larger than ${\cal O}({\rm nb})$, despite the large
value of $\Lambda$. This reflects (i) the strong $s$-channel
enhancement of the cross section, due to the small ALP width and (ii)
the large gluon content of the proton at LHC collision energies.

One can distinguish roughly three different kinematic regimes in
fig.~\ref{fig:xsec}. First, for $m_a \le 2 m_N$, the ALP in the
production diagram has to be off-shell, leading to only moderate cross
sections. We note that this is the case discussed in
\cite{deGiorgi:2022oks,Marcos:2024yfm}. Second, for $2 m_N \le m_a
\lsim 1$ TeV the ALP is produced on-shell and decays to two on-shell
HNLs. Because the ALP prefers to decay into heavy fermions, the
maximum of the cross section is always found for $m_a$ slightly above
$2 m_N$. Finally, for very heavy ALPs, masses roughly above
  $m_a=(1-2)$ TeV the cross section drops rapidly as a function of
the ALP mass, with $\sigma(pp \to a^{(*)} \to NN) \propto
1/(m_a\Lambda)^4$, i.e. effectively it behaves as a $d=8$ operator.

\begin{figure}[t]
    \centering
    \hspace{2cm}
    \includegraphics[scale=0.9]{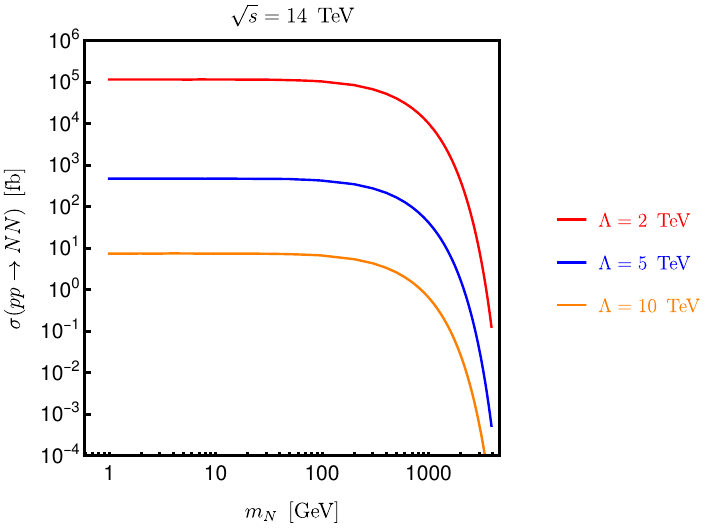}
    \caption{Cross section $\sigma(pp \to NN)$ in fb for $\sqrt{s}=14$
      TeV as a function of the HNL mass, $m_N$, for three different choices
      of the operator scale, $\Lambda$. The Wilson coefficient of the
      $d=7$ effective operator has been chosen as $c_{GN}=1$ in this
      example.}
    \label{fig:xsec2}
\end{figure}

Fig.~\ref{fig:xsec2} shows some example cross sections for the
process $pp \to NN$ for the $d=7$ operator given in eq.~\eqref{eq:d7},
choosing $c_{GN}=1$. We note that results for the choice $c_{{\tilde G}N}=1$
are very similar. Since the high-luminosity LHC should produce around
${\cal L}=3$ ab$^{-1}$ of luminosity, even for $\Lambda=10$ TeV one expects
between $(2-20)\times 10^3$ events for HNL masses between $(10-1000)$ GeV.

\subsection{Sensitivity estimates \label{subsec:res}}

Following the discussion of the cross section given above, we will
define three scenarios with finite ALP mass for our sensitivity
estimates. Scenario-I will use $m_a=5$ GeV; scenario-II uses $m_a=500$
GeV and scenario-III, $m_a=5$ TeV. This covers one example mass in
each of the three regimes shown in fig.~\ref{fig:xsec}. In addition,
we will show also results using the effective $d=7$ operator.

\begin{figure}[t]
    \centering
    \includegraphics[scale=0.42]{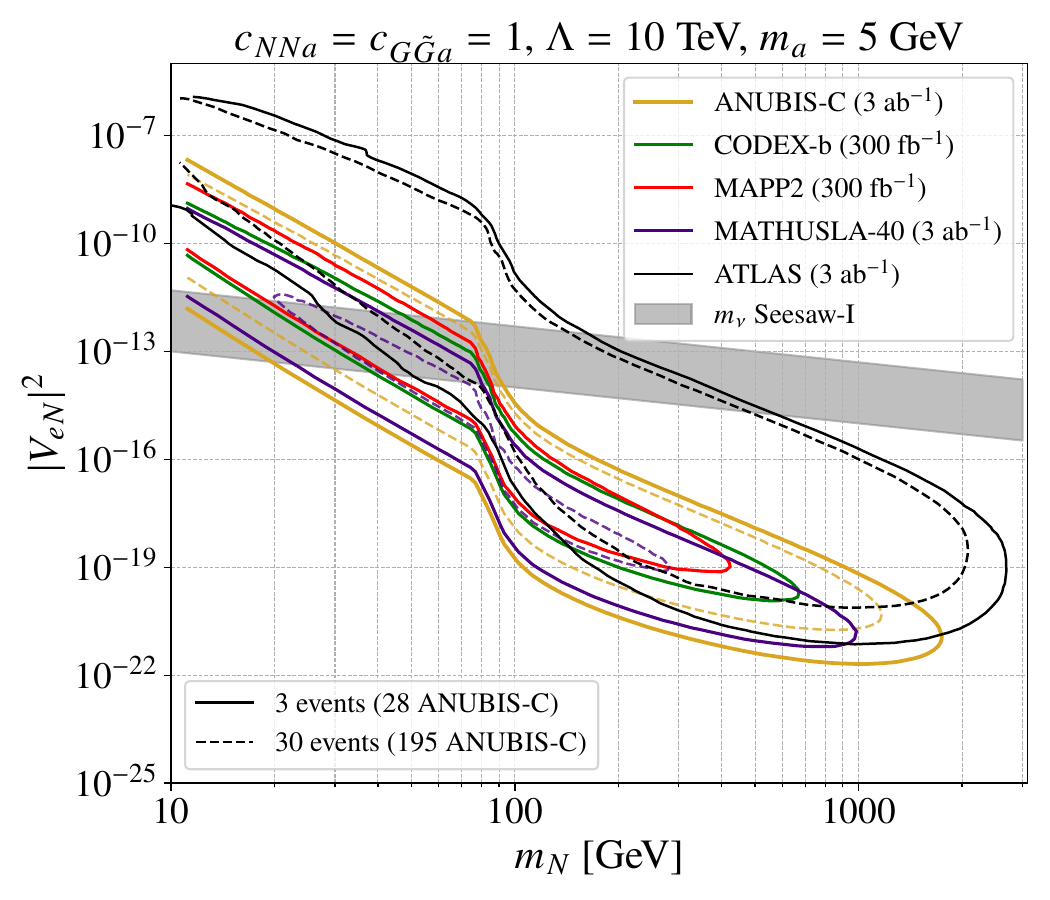}
    \includegraphics[scale=0.42]{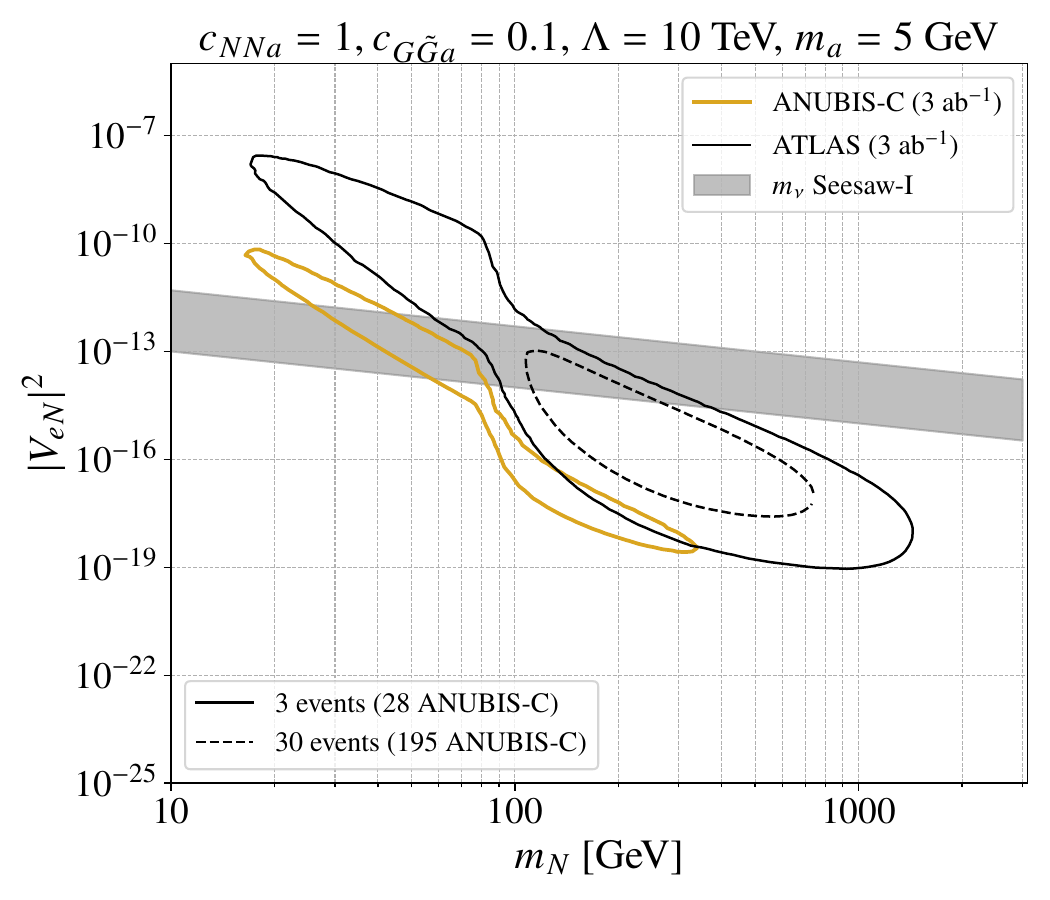}
    \includegraphics[scale=0.42]{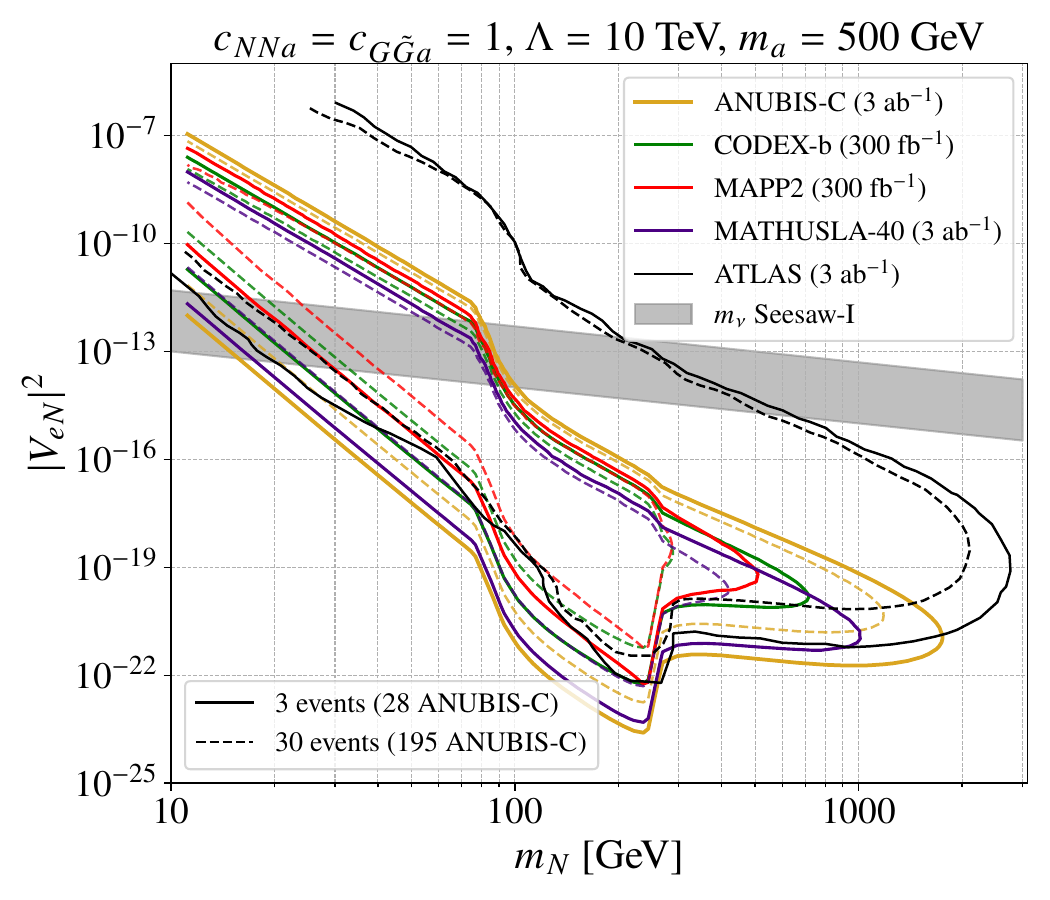}
    \includegraphics[scale=0.42]{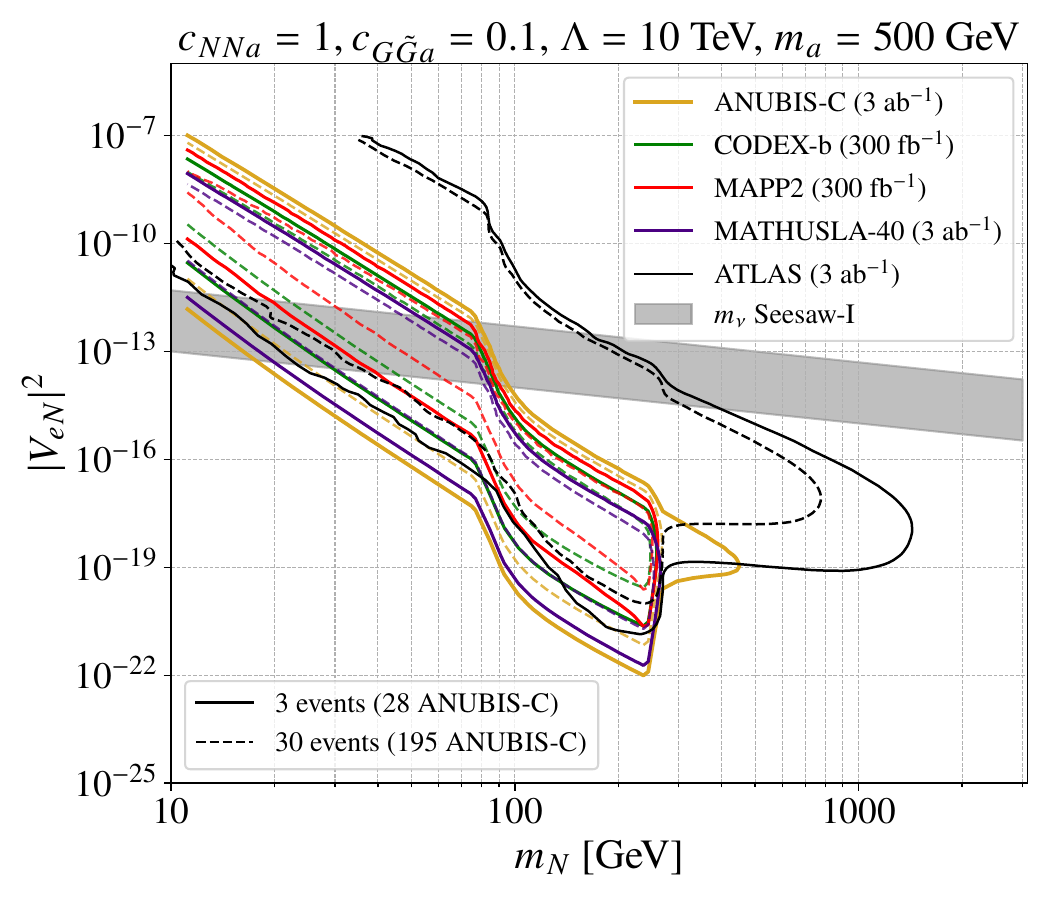}
    \caption{Sensitivity estimates for the high-luminosity LHC for
      HNL parameters for two different ALP masses, $m_a$ and two
      different combinations of Wilson coefficients at a fixed
      value of $\Lambda$. Solid (dashed) lines correspond to 3 (30) events for all experiments, except for ANUBIS-C, for which they represent 28 (195) event contours. For discussions, see text.}
    \label{fig:MixVmass}
\end{figure}

In fig.~\ref{fig:MixVmass}, we show sensitivity estimates for various
far detectors,\footnote{Only the transverse far detectors are
  sensitive to the scenarios considered here. We have checked that the
  forward detectors, namely FASER1 and FASER2, have no sensitivity.}
as well as ATLAS, in the plane ($|V_{eN}|^2,m_N$).The plots assume a
  fixed value of $\Lambda=10$ TeV in all cases and a collected LHC
  luminosity of ${\cal L}=3$ ab$^{-1}$ (300 fb$^{-1}$) for ATLAS,
  MATHUSLA-40 and ANUBIS-C (CODEX-b and MAPP2). Here, and in the
plots shown below, the full (dashed) contour lines for the sensitivity
of the different experiments are drawn for 3 (30) events.\footnote{We note in passing that the 30 event line for ATLAS is,
  of course, also the sensitivity limit for 3 events with 300
  fb$^{-1}$, roughly the current statistics and slightly less than the
  expected 500 fb$^{-1}$ for the sum of run-2 and run-3 of the
  LHC.}  The exception is ANUBIS-C, for which we use 28 (195) events
for the full (dashed) contours, as discussed at the beginning of this
section.  Recall, that 3 events correspond roughly to 95\% C.L. limits
for zero background.

The figure shows the results for two choices of the ALP mass,
corresponding to scenario-I (top panels) and scenario-II (bottom
panels). For both scenarios, we also use two combinations of the
Wilson coefficients: $(c_{G{\tilde G}a},c_{NNa}) =(1,1)$ in the left
plots and $(c_{G{\tilde G}a},c_{NNa})=(0.1,1)$ in the right plots.
The sensitivity range for $|V_{eN}|^2$ covers from $10^{-6}$ down to $10^{-23}$. For comparison, the grey band shows a rough expectation
for $|V_{eN}|^2$ for the type-I seesaw mechanism ($|V|^2 \sim
m_{\nu}/m_N$) using the range $m_{\nu} =[1,50]$ meV.

For the parameter choices shown, the different detectors cover the seesaw
band in a large range of masses. One can see that the far detectors,
in general, tend to probe smaller values of mixing than ATLAS. This
simply reflects the larger distance to the IP. Despite the fact that
for ANUBIS-C we have to expect non-zero backgrounds and thus draw the
sensitivity line at 28 events, ANUBIS-C probes the smallest mixing of
all experiments.  We note in passing that the MATHUSLA-40 3-event line
is nearly identical to the original MATHUSLA configuration 60-event
line, i.e. a loss in sensitivity of around a factor 20 in number of
events, see also the discussion in appendix \ref{sec:appB}.

One also notes that not all simulated experiments show sensitivity in
all plots. In particular, comparing the plots for $c_{G{\tilde G}a}=0.1$
to those for $c_{G{\tilde G}a}=1$ one sees that while similar ranges
of parameters are probed in both cases, there is a loss of sensitivity
due to the reduced cross section for $c_{G{\tilde G}a}=0.1$.  One
notices that for $m_a=5$ GeV and $c_{G{\tilde G}a}=0.1$ (upper
  right plot) only ATLAS and ANUBIS-C reach the necessary event
numbers to provide constraints. This is very different for the case
with the largest cross section, with $m_a=500$ GeV and $c_{G{\tilde G}a}=1$
(bottom left plot), where many of the proposed far
detectors show good sensitivity. On the other hand, the plots make it
clear that for $c_{G{\tilde G}a} \ll 0.1$ the cross sections are 
too low to produce a sufficient number of events and all sensitivity
will be lost.

\begin{figure}[t]
    \centering
    \includegraphics[scale=0.42]{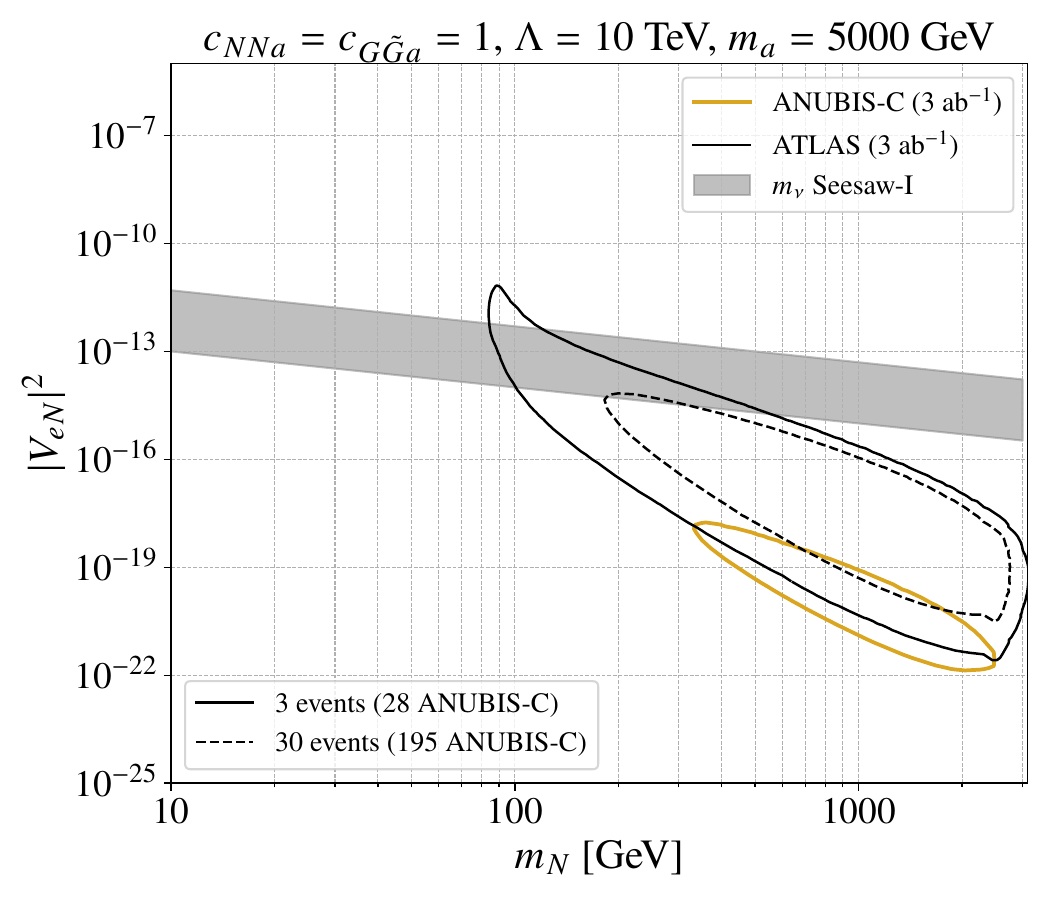}
    \includegraphics[scale=0.42]{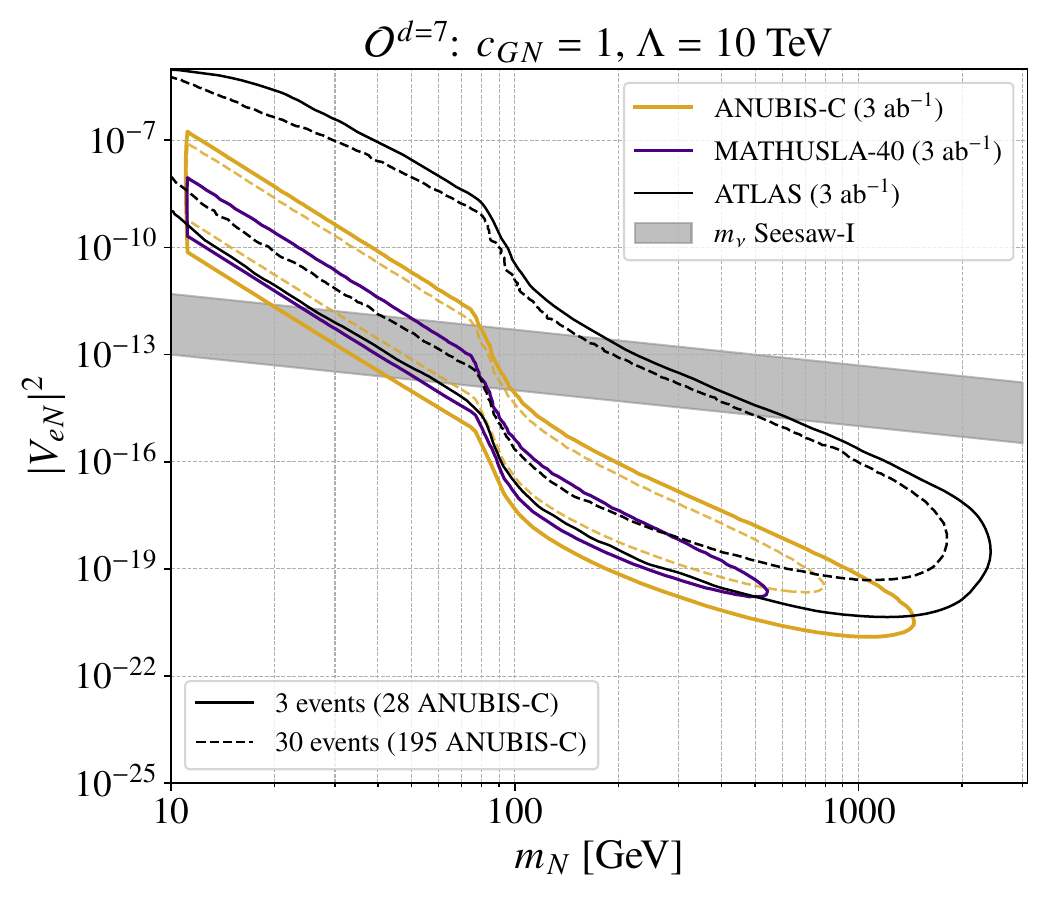}
    \caption{Sensitivity estimates for the high-luminosity LHC for
      HNL parameters for two different cases: To the left,
      a calculation putting $m_a=5$ TeV; to the right, using the
      $d=7$ operator with $\Lambda = 10$ TeV. Solid (dashed) lines correspond to 3 (30) events, except for ANUBIS-C, for which they correspond to 28 (195) events.}
    \label{fig:MixVmass2}
\end{figure}

In fig.~\ref{fig:MixVmass2}, we show the result for $m_a=5$ TeV and
$c_{G{\tilde G}a}=1$, corresponding to scenario-III, and compare this
with a calculation using the $d=7$ operator with $c_{GN}=1$ and
$\Lambda=10$ TeV. For $m_a=5$ TeV the calculation using the ALP is
already quite close to an effective theory with a $d=8$ operator, as
has been discussed above. Because the branching ratio of the ALP to
two right-handed neutrinos is proportional to $m_N^2$, sensitivity is
found only for large values of $m_N$. For the $d=7$ operator, even at
$\Lambda=10$ TeV, sensitivity exists over the whole range of
right-handed neutrino masses. The width of the sensitivity region in
$|V_{eN}|^2$ is not as wide for the $d=7$ operator as for the case of
small ALP masses, discussed above. Again this reflects the smaller
cross section for the $d=7$ operator, relative to the on-shell light
ALP case, compare figs.~\ref{fig:xsec} and \ref{fig:xsec2}.

\begin{figure}[t]
    \centering
    \includegraphics[scale=0.42]{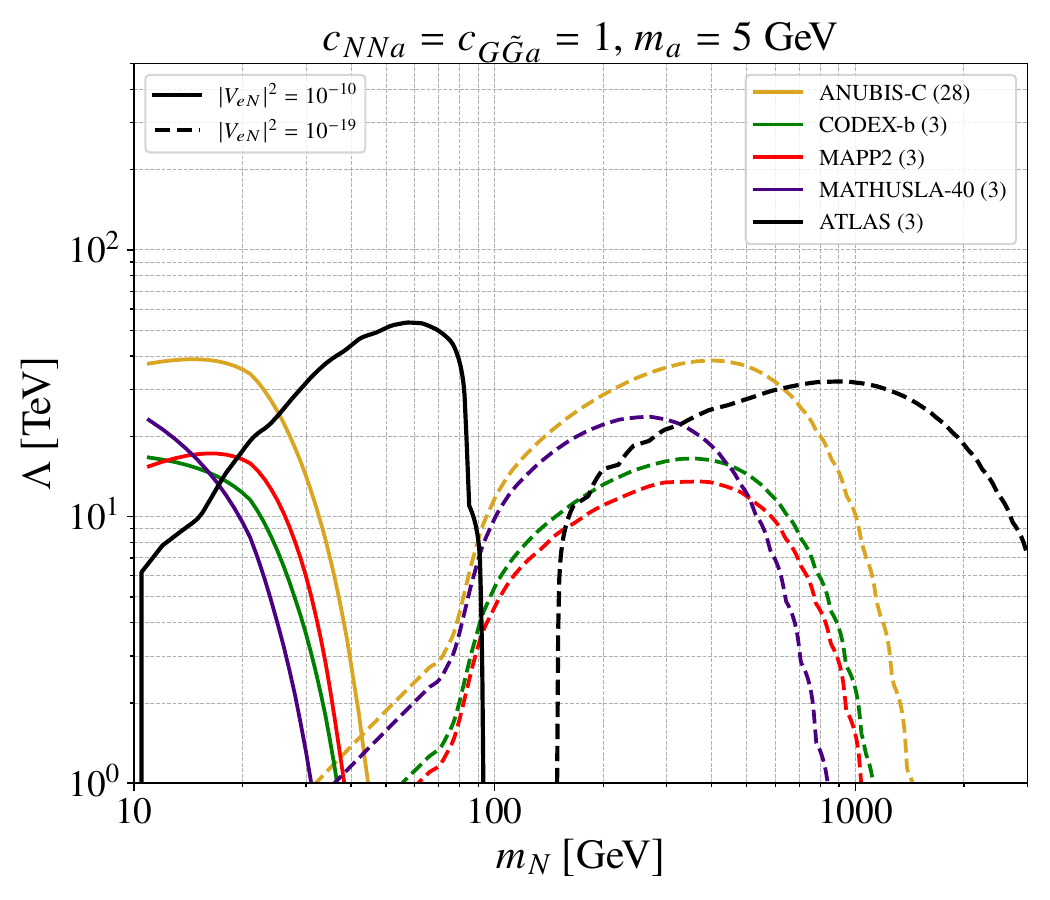}
    \includegraphics[scale=0.42]{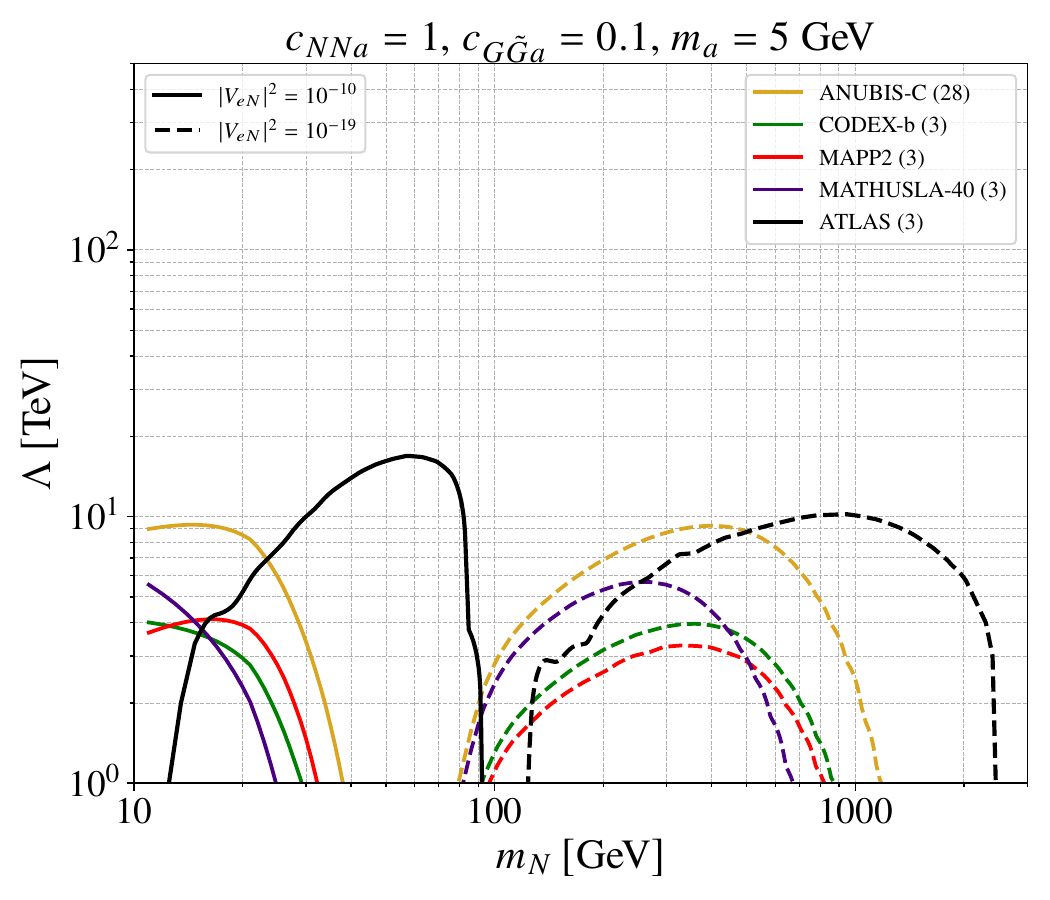}
    \includegraphics[scale=0.42]{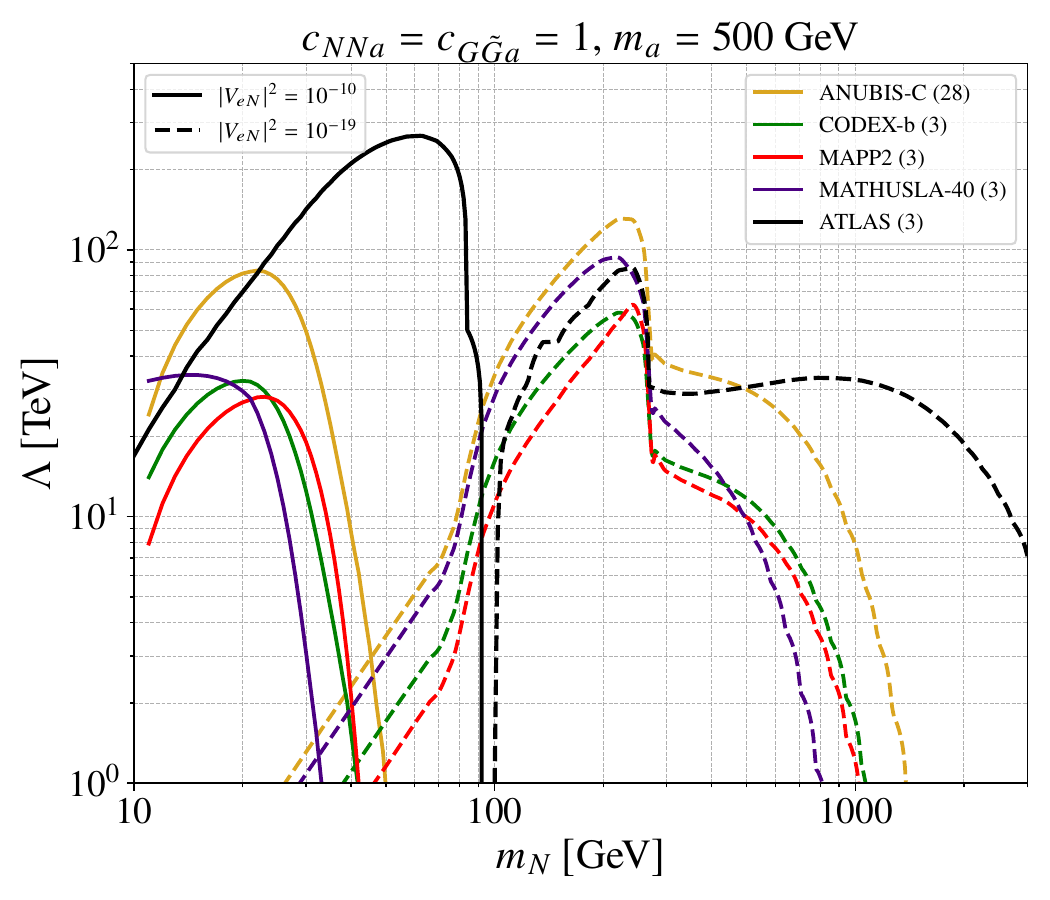}
    \includegraphics[scale=0.42]{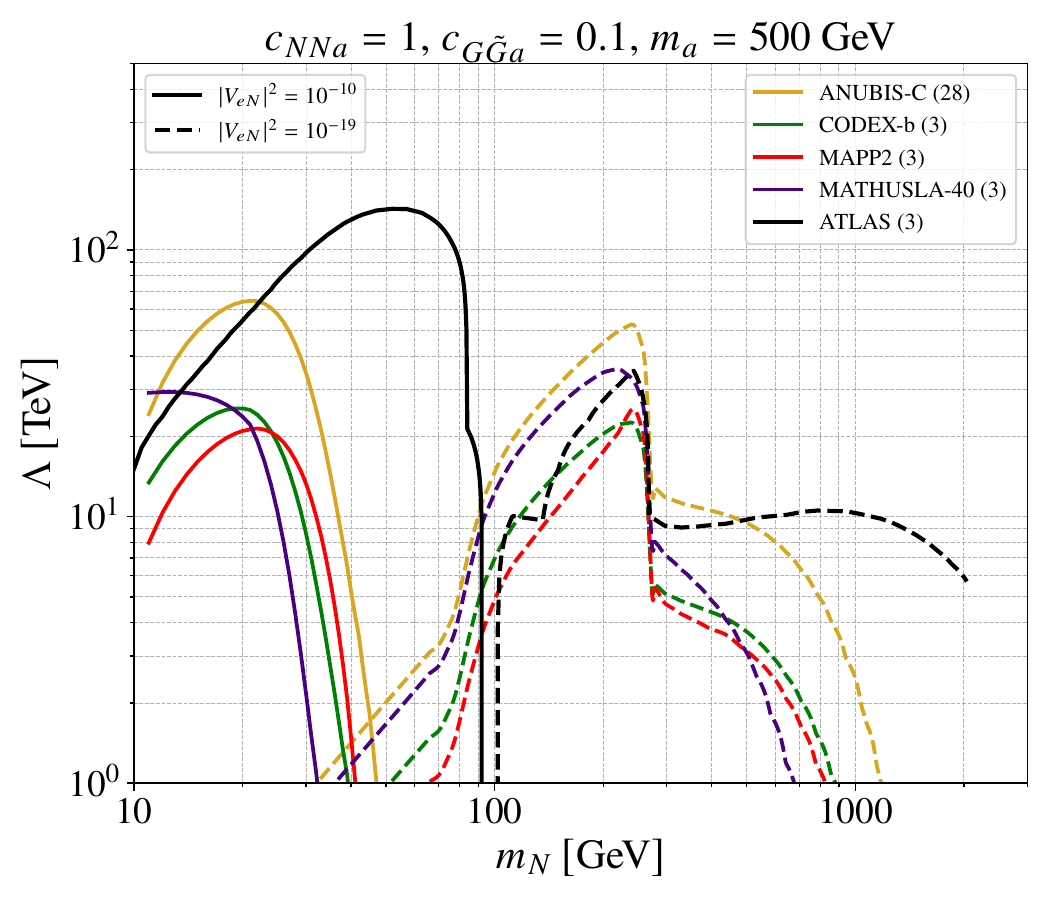}
    \caption{Sensitivity estimates for the high-luminosity LHC for
      $\Lambda$ for two different ALP masses, $m_a$ and two different
      combinations of Wilson coefficients. Solid and dashed lines
      correspond to two specific values of the mixing squared
      parameter. Here, we use 3 events as the sensitivity line for
        all experiments, except for ANUBIS-C, for which we use 28
        events. }
    \label{fig:LamVmass}
\end{figure}

It is also interesting to ask, what is the maximal value of $\Lambda$,
to which the different experiments will be sensitive in the current
model.  In fig.~\ref{fig:LamVmass}, we show results for the reach in
$\Lambda$ for the same ALP masses and couplings as in
fig.~\ref{fig:MixVmass} for two choices of $|V_{eN}|^2$, one
relatively large and one very small. As shown, ATLAS should have the
largest sensitivity in $\Lambda$, followed by ANUBIS-C. For the most
optimistic choice of parameters, ATLAS will be sensitive up to roughly
$\Lambda \simeq 300$ TeV, and in the most pessimistic case
shown, roughly $\Lambda \simeq 10$ TeV. Also the far detectors have
reaches in the range of several $10$ TeV, with ANUBIS-C reaching $\Lambda \ge 100$ TeV in the most optimistic case.

 \begin{figure}[t]
    \centering
    \includegraphics[scale=0.42]{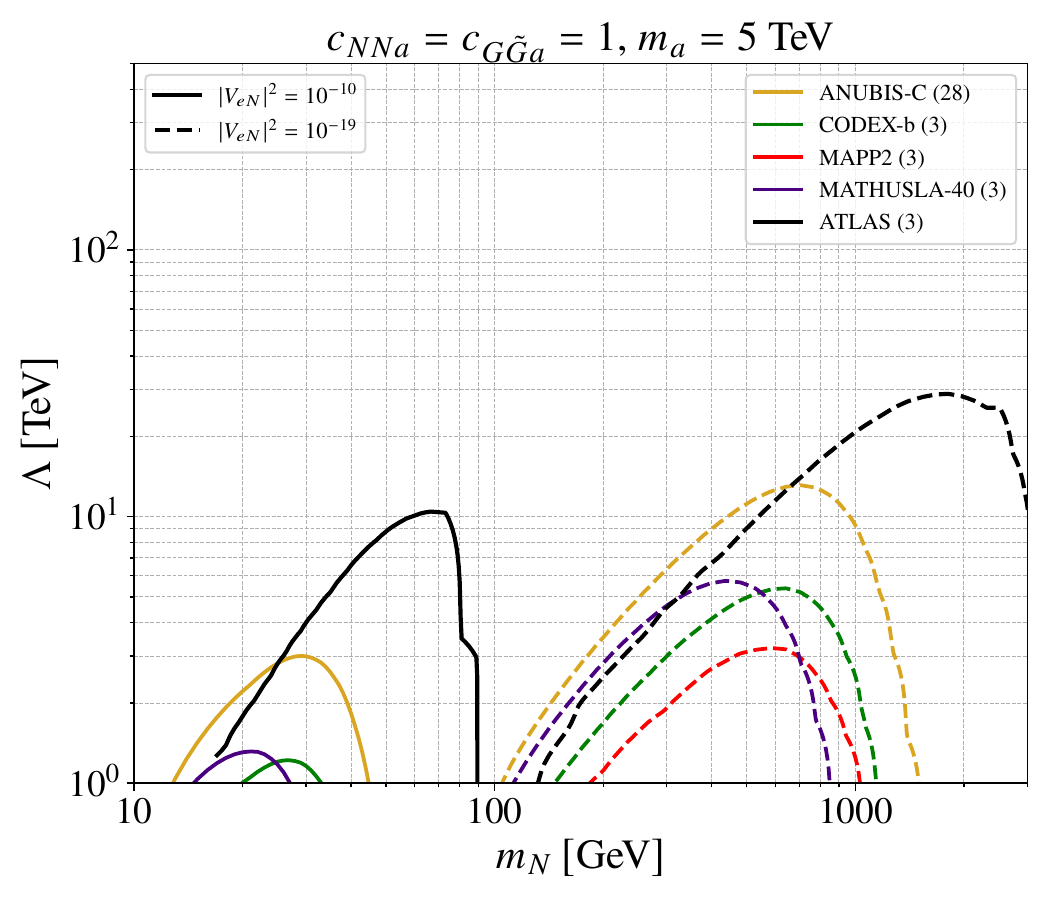}
    \includegraphics[scale=0.42]{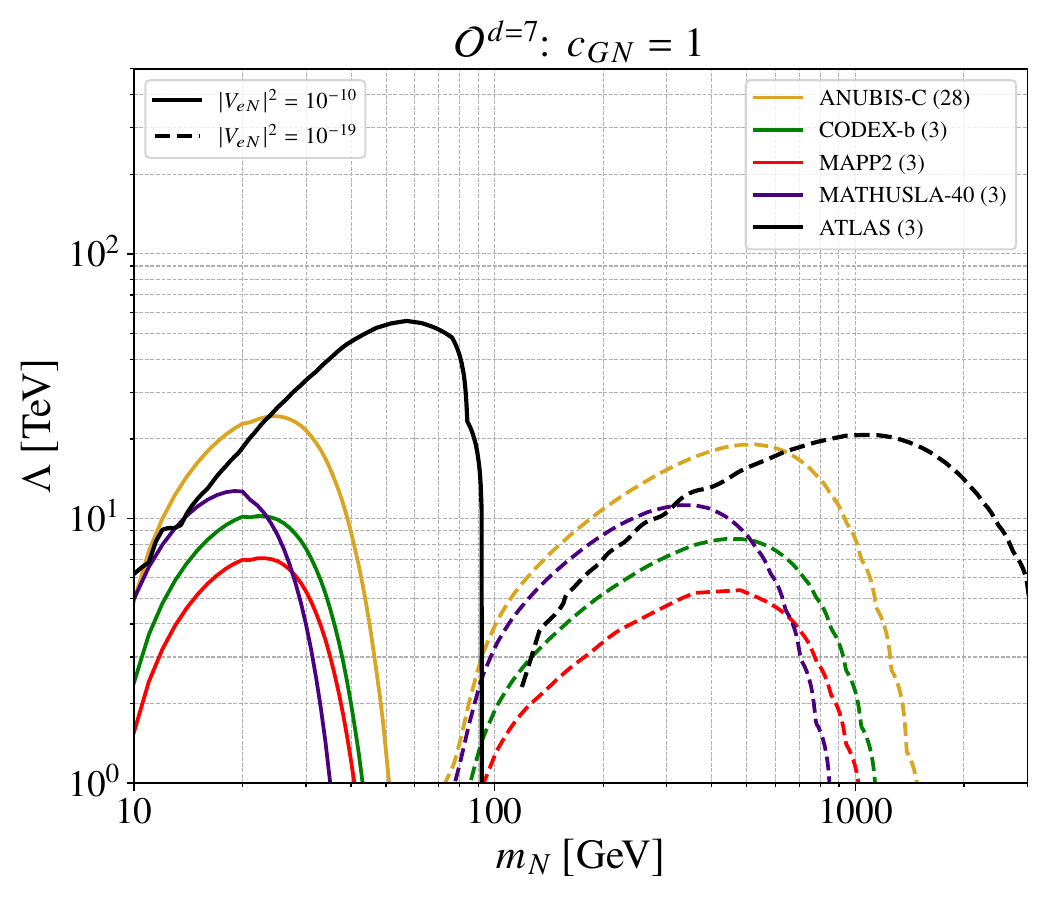}
    \caption{Sensitivity estimates for the high-luminosity LHC for
      $\Lambda$ for two different cases: To the left,
      a calculation putting $m_a=5$ TeV; to the right, using the
      $d=7$ operator. Solid and dashed lines
      correspond to two specific values of the mixing squared
      parameter. Here, we use 3 events as the sensitivity line for
      all experiments, except for ANUBIS-C, for which we use 28
      events.}
    \label{fig:LamVmass2}
\end{figure}

Finally, we show in fig.~\ref{fig:LamVmass2} the results in the
($\Lambda, m_N$) plane for the same scenarios as in
fig.~\ref{fig:MixVmass2}. The plot on the left shows already a much
reduced reach in $\Lambda$, compared to the lighter ALP masses used
in fig.~\ref{fig:LamVmass}. For the $d=7$ operator again ATLAS
and ANUBIS-C show the best sensitivities of up to 50 TeV and 20 TeV
for $|V_{eN}|^2=10^{-10}$.

In summary, HNLs coupled to gluons -- either from a $d=7$ operator
or through the ALP portal -- have large production cross sections
at the LHC. Thus, this scenario provides exceptional sensitivity
to HNL parameters, with values of the mixing $|V_{eN}|^2$ below
even the seesaw line accessible in many cases.

% !TEX root = ../NRALP.tex
\section{Conclusions\label{sec:cncl}}

In this paper we have discussed the sensitivity of future LHC
searches for long-lived HNLs in two different theoretical setups. In
the first model, we add HNLs and an ALP with an arbitrary mass. HNLs
are produced at the LHC from the couplings of the ALP to gluons and
HNLs. In the second variant, we use a $d=7$ $N_R$SMEFT operator to
couple HNLs to gluons. For both model variants we have calculated
sensitivity estimates for the high-luminosity LHC.

For the ALP model, in the most favourable cases HNLs with mixings as
small as $|V|^2 = 10^{-24}$ can be probed in a large range of HNL
masses. We have also shown that for Wilson coefficients of order
unity, future LHC data will be sensitive to $\Lambda$ scales for the
effective coupling of ALPs to gluon and HNLs up to $\Lambda=300$ TeV
in the best case. For the $d=7$ operator, due to lower production
cross sections, the sensitivity is expected to be weaker, but still
in the range of (20-50) TeV for ANUBIS-C and ATLAS.

Finally, let us briefly mention that in this paper we have
concentrated exclusively on the couplings of HNLs to gluons. However,
both, the ALP Lagrangian as well as the full set of $N_R$SMEFT
operators, include terms that couple HNLs to the $B^{\mu\nu}$ and
$W^{\mu\nu}$ field strength tensors. Thus, one could probe HNLs also
in associaton with vector-boson fusion type diagrams. The cross
section for this production mode will be lower than the gluonic ones
considered in this paper, but the events will have additional forward
jets. This could be used as a tag to identify these particular
operators/couplings and, despite lower cross sections, there should be
ample parameter space to be explored.

\bigskip
\bigskip

\centerline{\bf Acknowledgements}

\bigskip

The authors would like to thank Zeren Simon Wang and Giovanna Cottin
for many discussions on long-lived particles over the past years. We
also thank Zeren Simon Wang for help with his code DDC
\cite{Domingo:2023dew} and Giovanna Cottin for a private copy of
her code for the simulation of displaced vertices in ATLAS.  We
acknowledge support by Spanish grants PID2023-147306NB-I00 and
CEX2023-001292-S (MCIU/AEI/10.13039/501100011033), as well as
CIPROM/2021/054 (Generalitat Valenciana).  R.~B. is supported by the
grant ACIF/2021/052 (Generalitat Valenciana).  C.~H. is funded by the
Generalitat Valenciana under Plan Gen-T via CDEIGENT grant
No. CIDEIG/2022/16. A.~M. is funded by the Generalitat Valenciana via
grant No. CIDEXG/2022/20.

\bigskip

\appendix
% !TEX root = ../NRALP.tex
\section{Dijet constraints on $c_{G{\tilde G}a}$\label{sec:app}}

If ALPs with couplings to gluons exist in the mass range
  $m_a=[2,2000]$ GeV, they will be produced with large rates at the
  LHC.  This straight-forward observation forms the basic motivation
  of our current paper. Limits on the coupling $c_{G{\tilde
      G}a}/\Lambda$ are therefore the most important constraints on
  models with ALPs for our current work. To the best of our knowledge,
  the currently best limits on $c_{G{\tilde G}a}/\Lambda$ can be
  derived from a reinterpretation of dijet searches at the LHC.

  Decays of the ALPs will contribute to dijet (and trijet
\cite{Ghebretinsaea:2022djg}) events. Therefore, we briefly discuss how
searches for dijet resonances at LHC can be used to derive upper
limits on $c_{G{\tilde G}a}/\Lambda$ with current data. For this
purpose, we have made a reinterpretation of upper limits on the BSM
dijet cross sections from two searches, one by CMS \cite{CMS:2019gwf}
and one by ATLAS \cite{ATLAS:2018qto}. These searches cover different
invariant dijet mass regions. The CMS search \cite{CMS:2019gwf} is
sensitive at the largest dijet invariant mass, $m_{jj}=[1.8,8.8]$ TeV,
while ATLAS \cite{ATLAS:2018qto} covers the range $m_{jj}=[0.45,1.8]$
TeV.

\begin{figure}[t]
    \centering
    \includegraphics[scale=0.9]{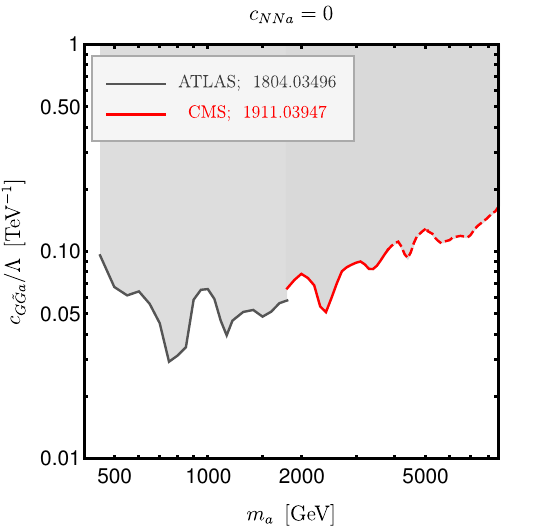}
    \caption{Limits on $c_{G{\tilde G}a}/\Lambda$ in [TeV$^{-1}$] as a
      function of ALP mass, $m_a$, from a reinterpretation of two
      experimental dijet searches \cite{ATLAS:2018qto,CMS:2019gwf}.
      Note the double logarithmic scale. For explanation see text.}
    \label{fig:LimcGG}
\end{figure}

Fig.~\ref{fig:LimcGG} shows our estimated upper limits on
$c_{G{\tilde G}a}/\Lambda$ based on this data. These limits suffer
from a number of uncertainties, which we need to discuss in some detail,
as well as from some theory assumptions, on which we comment below.

First of all, the CMS collaboration assumes in most parts of their
analysis \cite{CMS:2019gwf} that the width of the BSM resonance is
negligible compared to the experimental dijet mass resolution. The
decay width of the ALP, however, is proportional to
$(c_{G{\tilde G}a}/\Lambda)^2$.\footnote{If we assume all other Wilson
coefficients are zero, as it is done in fig.~\ref{fig:LimcGG}.}  For a
value of $(c_{G{\tilde G}a}/\Lambda) \simeq 0.1$ TeV$^{-1}$ the ALP
width becomes larger than 10\% of its mass for $m_a \ge 4 $ TeV, thus
violating the small width assumption.  Section 6.2 of
\cite{CMS:2019gwf} discusses how limits would change for broad BSM
resonances. However, that discussion is limited to spin-1 and spin-2
resonances, but does not cover spin-0 states. Relative to the results
on narrow resonances, shown in fig.~6 of \cite{CMS:2019gwf}, limits
deteriorate strongly as a function of both, resonance mass and
width. While for spin-2 resonances, fig.~10 of that paper shows
limits up to the largest masses, for spin-1 resonances the
collaboration shows no limits above $m_{jj}=6$ TeV. Since no results
for spin-0 resonances are shown, we have no way to take into account
how the ALP limits are changed in the high mass regime. Instead, we
simply show the derived limit for $c_{G{\tilde G}a}/\Lambda$, assuming
a small width, as function of the ALP mass, $m_a$. For the cases where
at the derived limit $\Gamma_{a}/m_a \le 5$\%, we show the limit as
full line, whereas for the mass region where the ALP width is larger
than the 5\% at the derived limit for $c_{G{\tilde G}a}/\Lambda$, we
show a dashed line. We consider limits shown as dashed lines as not
reliable at the moment.  However, because the ALP width is
proportional to $(c_{G{\tilde G}a}/\Lambda)^2$, with future, improved
limits on the BSM dijet cross section, also this large mass window
will be probed.

Next, CMS divides the search into quark-quark, quark-gluon and
gluon-gluon jets. From fig.~6 of \cite{CMS:2019gwf}, one can see that
limits are roughly similar but not identical, depending on whether
quarks or gluon jets are assumed in the analysis. For masses in the
range of roughly $m_{jj} \simeq [1.8,6]$ TeV, limits on the gluon jets
are typically a factor of two worse than limits on quark jets. For the
largest invariant masses, limits on gluon jets start to deteriorate and
can be up to one order of magnitude worse than limits on quark
jets. In fig.~\ref{fig:LimcGG}, we use the limits on gluon jets in
the CMS mass window.

For lower invariant masses, SM contributions to the dijet
rate at the LHC have too high rates for the CMS search
\cite{CMS:2019gwf} to be effective. ATLAS \cite{ATLAS:2018qto}
therefore pesented a dedicated analysis, recording only the event
information calculated by the jet trigger algorithms, thus lowering
the trigger rates to an acceptable level. The ATLAS analysis is
strictly speaking only valid for quark jets, initiated by a spin-1 BSM
resonance. No results are presented for gluon jets, nor for spin-0
resonances. To be conservative, our reinterpretation of
\cite{ATLAS:2018qto} therefore includes a factor of two on the limit
on the BSM cross section, motivated by the results from CMS
\cite{CMS:2019gwf}, discussed above, that limits on gluon jets will be
worse than limits on quark jets roughly by this factor.  However, the
CMS result is based on a different $m_{jj}$ window and thus this
estimate has to be taken with a grain of salt.  Only the experimental
collaborations can derive more accurate bounds, including spin-0
resonances in their analysis in future searches.

Finally, the results shown in fig.~\ref{fig:LimcGG} also depend,
quite strongly, on the assumption that all other Wilson coefficients
vanish identically, $\forall c_i=0$. If other Wilson coefficients are
non-zero, the branching ratio for Br($a\to jj$) is no longer 100\%
and the limits shown in the figure will deteriorate roughly by $\sim
1/\sqrt{{\rm Br}(a \to jj)}$. In this paper, we are mostly interested
in the production of HNLs, thus always assume $c_{NNa}$ is
non-zero. The ALP width to fermions, however, is proportional to the
fermion mass squared, thus Br($a\to jj$) becomes a function of $m_N$
and $c_{NNa}$, if $c_{NNa}$ is non-zero. Br($a\to NN$) is shown in
fig. \ref{fig:BrNN} for different choices of couplings and masses
and the limits shown in fig.~\ref{fig:LimcGG} will be weaker for
those cases by the corresponding reduction in Br($a\to jj$).

\section{Impact of the design updates for MATHUSLA and
  ANUBIS\label{sec:appB}}

In this appendix, we will compare sensitivity estimates for different
versions of two of the far detectors, MATHUSLA and ANUBIS, that both
have been updated recently. The Displaced Decay Counter (DDC), which
we are using in our calculations, has recently been updated to include
the latest designs of these two experiments, but keeps also the
geometry of the earlier versions \cite{Domingo:2023dew}.

We will start the discussion with MATHUSLA. MATHUSLA was first
discussed in \cite{Chou:2016lxi}, a letter of intent was later
published in \cite{MATHUSLA:2018bqv}. In these early versions MATHUSLA
was proposed as a massive detector of dimensions $200\times 200 \times
20$ m$^3$ to be placed above ATLAS. An update published in
\cite{MATHUSLA:2020uve} discussed to have MATHUSLA built above
CMS. The detector was thought to be moved closer to the IP, allowing
to have very similar sensitivity than the original version, despite
being nearly four times smaller, i.e.  $100\times 100 \times 25$
m$^3$. This version was originally implemented in DDC
\cite{Domingo:2023dew}. However, in the design report, published very
recently in \cite{MATHUSLA:2025zyt}, the dimensions of the detector
were reduce to a much more moderate $40\times 40\times 16$ m$^3$.  We
refer to this design as the MATHUSLA-40.

\begin{figure}[t]
    \centering
    \begin{subfigure}[b]{0.49\textwidth}
        \centering
        \includegraphics[width=\textwidth]{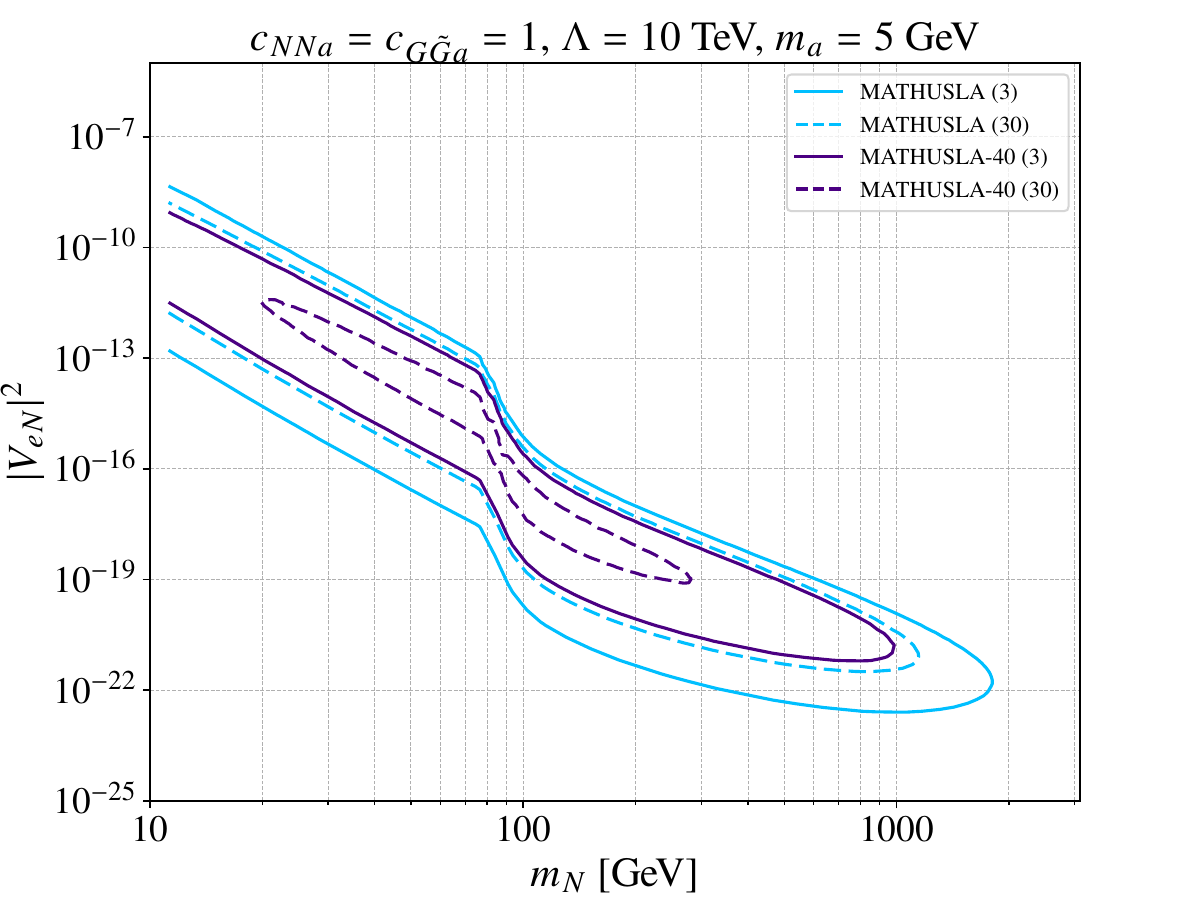}
    \end{subfigure}
    \hfill
    \begin{subfigure}[b]{0.49\textwidth}
        \centering
        \includegraphics[width=\textwidth]{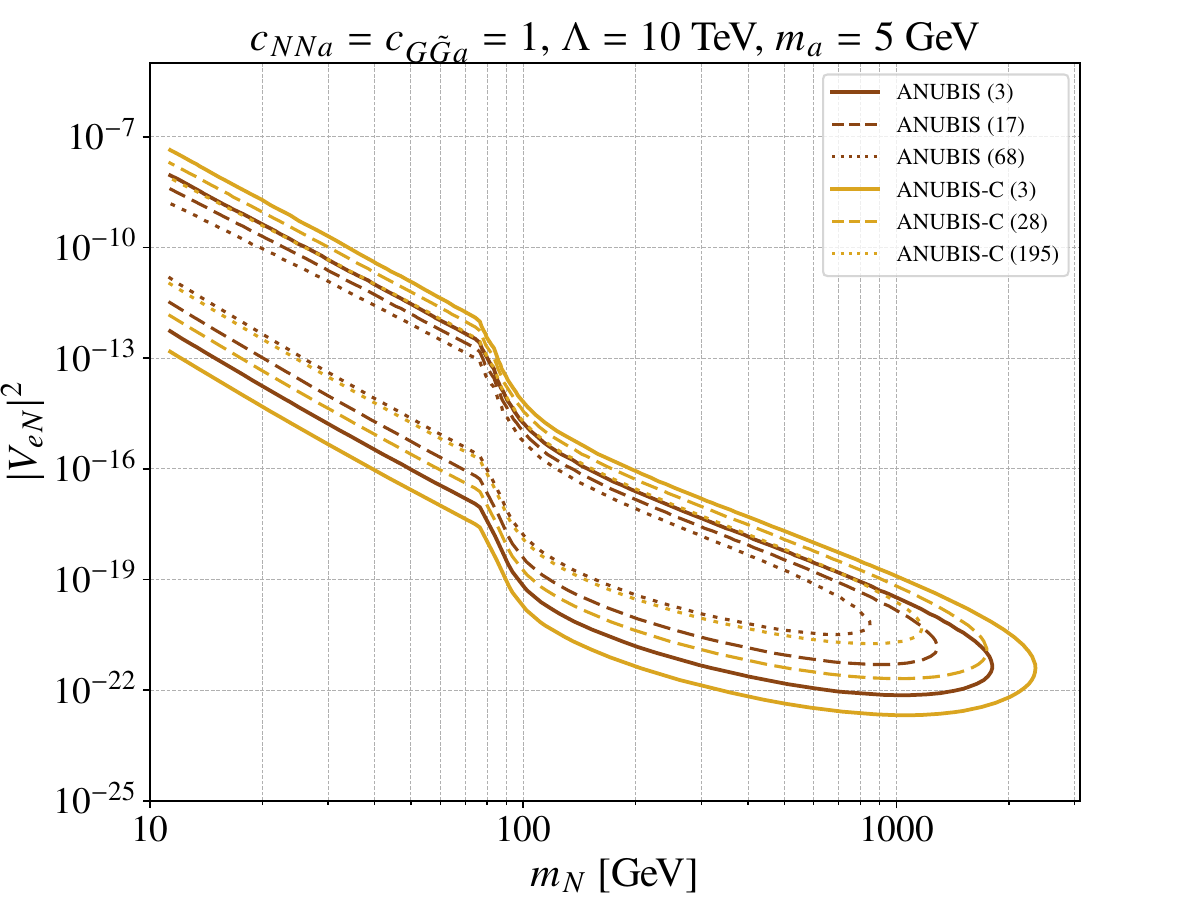}
    \end{subfigure}
    \caption{Comparison of sensitivity between two design versions of
      MATHUSLA (left), and two designs for ANUBIS (right) for one
      particular scenario of our ALP+HNL setup, as indicated in the plots.}
    \label{fig:DsgnComp}
\end{figure}

In fig.~\ref{fig:DsgnComp}, to the left, we show the estimated
sensitivities for both MATHUSLA and MATHUSLA-40, for one particular
scenario in our model setup, as indicated in the plots.  We have
checked that the sensitivity region for 60 events for the larger
MATHUSLA design is nearly identical to the 3 event line of MATHUSLA-40
in this scenario. That is, the loss of sensitivity is similar, but
slightly larger, than the ratio of the volumes of the two designs. We
have made the same comparison also for the other parameter choices of
our model, discussed in the main text. We find in all cases similar
reduction factors, except for the largest HNL masses, where the rapid
decrease of the production cross sections, see figs. \ref{fig:xsec}
and \ref{fig:xsec2}, leads also to a notably reduced mass reach
of the smaller MATHUSLA-40.

The original ANUBIS proposal \cite{Bauer:2019vqk} imagined to install
the detector in a vertical service shaft above ATLAS. The updated
configuration, currently discussed by the collaboration
\cite{Brandt:2025fdj}, plans to install the detector components
directly onto the ceiling of the ATLAS cavern. This geometry is
referred to as ANUBIS-C throughout this paper. ANUBIS-C is
considerably closer to the interaction point (IP) than the original
proposal, thus it covers a larger solid angle and its peak sensitivity
is shifted towards smaller decay lengths.

A major concern for all LLP searches is the suppression of
backgrounds. This is particularly true for ANUBIS, which is relatively
close to the ATLAS IP and with no other passive shielding than the
ATLAS detector itself. Consequently, the ANUBIS collaboration in their
last two publications \cite{Brandt:2025fdj,ANUBIS:2025sgg} has
paid special attention to discuss backgrounds. It is planned that
ANUBIS will be fully integrated with the ATLAS detector, such that
both experiments can serve as a trigger for the other. According to
\cite{ANUBIS:2025sgg} this active veto, together with the ATLAS
  calorimeter acting as a passive veto, will allow to reduce
  backgrounds in ANUBIS to at most $\sim 182.4\pm 12.2$ events in $3$
  ab$^{-1}$ of luminosity. We mention, however, that ANUBIS
  \cite{ANUBIS:2025sgg} also gives a more optimistic background
  estimate of $51.3\pm 3.2$ events, when using the results from a
  recent ATLAS paper \cite{ATLAS:2025pak} that uses machine learning
  techniques to reduce backgrounds. Following ANUBIS
  \cite{ANUBIS:2025sgg}, we decided not to use this more optimistic
  calculation in our sensitivity estimates.

For our estimates of the senstivity of ANUBIS, we thus decided to use
$2 \times \sqrt{195} \simeq 28$ events, as the expected 95\%
contour for all the figures in the main text. This rather simplistic
treatment relies on the assumption that the number of background
events is well understood, which seems to be rather
  optimistic. Here, in fig.~\ref{fig:DsgnComp}, to the right, we
compare the original ANUBIS proposal \cite{Bauer:2019vqk} to ANUBIS-C
\cite{Brandt:2025fdj}. Also for the original ANUBIS configuration,
  called ``ANUBIS shaft'' in \cite{ANUBIS:2025sgg}, the latest ANUBIS
  paper provides a background estimate of $63.7 \pm 4.3$ events.
  To compare the old versus the new configuration of ANUBIS we therefore
  show in fig.~\ref{fig:DsgnComp} three lines for each design. For ANUBIS
  we use (3, 17, 68) events, while for ANUBIS-C we use (3, 28, 195) events.
  With this choice, one can see the effects of both, the change of design
  as well as the reduction of sensitivity for different background
  estimates.

As can be seen, ANUBIS-C is more sensitive than ANUBIS, for the same
number of events. This is expected from the larger effective volume
covered by ANUBIS-C relative to ANUBIS. It is also clear that ANUBIS-C
for 28 events will have very similar sensitivity as ANUBIS with 3
events. However, the region probed by ANUBIS-C is shifted slightly
towards larger values of mixing, when compared to ANUBIS. Again, this
simply reflects the fact that the ANUBIS-C design is closer to the IP
than ANUBIS. Recall that larger values of mixing imply smaller decay
lengths. Again, we have checked that the relative merits of ANUBIS and
ANUBIS-C are qualitatively similar also for all other scenarios we
study in this paper.

% Bibliography
\bibliographystyle{JHEP}
\bibliography{RefsNRALP.bib}

\end{document}